\documentstyle[epsf]{mn}

\title[]{{\sl XMM-Newton} observations of the asynchronous polar BY Cam
\thanks{Based on observations obtained with XMM-Newton, an ESA science
mission with instruments and contributions directly funded by ESA
Member States and the USA (NASA).}}

\author[G. Ramsay and M. Cropper]{Gavin Ramsay and Mark Cropper\\
Mullard Space Science Lab, University College London,
Holmbury St. Mary, Dorking, Surrey, RH5 6NT, UK\\}

\begin{document}
\outer\def\gtae {$\buildrel {\lower3pt\hbox{$>$}} \over 
{\lower2pt\hbox{$\sim$}} $}
\outer\def\ltae {$\buildrel {\lower3pt\hbox{$<$}} \over 
{\lower2pt\hbox{$\sim$}} $}
\newcommand{\ergscm} {ergs s$^{-1}$ cm$^{-2}$}
\newcommand{\ergss} {ergs s$^{-1}$}
\newcommand{\ergsd} {ergs s$^{-1}$ $d^{2}_{100}$}
\newcommand{\pcmsq} {cm$^{-2}$}
\newcommand{\ros} {\sl ROSAT}
\newcommand{\exo} {\sl EXOSAT}
\def\rchi{{${\chi}_{\nu}^{2}$}}
\newcommand{\xmm} {\sl XMM-Newton}
\newcommand{\Msun} {$M_{\odot}$}
\newcommand{\Mwd} {$M_{wd}$}
\def\Mdot{\hbox{$\dot M$}}
\def\mdot{\hbox{$\dot m$}}

\maketitle

\begin{abstract}
We report observations of the asynchronous polar BY Cam made using
{\xmm}. We find evidence for two accretion regions which have
significantly different spectra. In both regions we find evidence for
hard X-ray emission from the post-shock flow while we observe a
distinct soft X-ray blackbody component in only one. We detect two
emission lines in the RGS detectors which we attribute to Nitrogen NVI
and NVII: the first time that Nitrogen lines have been detected in an
X-ray spectrum of a cataclysmic variable. In the time interval where
we observe short timescale variability we find the hard X-ray light
curve is correlated with the softest energies in the sense that the
hard X-rays trail the soft X-rays and are anti-correlated. We suggest
that there are a small number of dense blobs of material which impact
the white dwarf without forming a shock and release an amount of
optically thick material which obscures hard X-rays for a short
duration.
\end{abstract}


\section{Introduction}

Polars are close binaries in which the primary star (a white dwarf)
accretes material from a red dwarf secondary star. The strength of the
magnetic field of the white dwarf (B$\sim$10--200\,MG) is sufficient
to synchronize its spin period with that of the binary orbital
period. The accretion flow is controlled over large distances by the
white dwarf's magnetic field. In most of these systems the magnetic
field is locked into a single orientation with respect to the
secondary star.

There are, however, four polars which are known not to be exactly
synchronous: the spin period of the white dwarf differs from the
orbital period by a few percent. It is expected that these systems are
only temporarily out of phase-lock (due to eg a nova explosion) and
will regain synchronization within several hundred years (Schmidt \&
Stockman 1991). One of the brightest of these asynchronous polars is
BY Cam. It is bright in X-rays and has an binary orbital period of
201.258 mins and a white dwarf with a spin period of 199.330 mins
(Mason et al 1998). This results in a spin-orbit beat period of 14.5
days.

BY Cam has been observed with various X-ray telescopes. Indeed, it was
discovered using the X-ray satellite {\sl Uhuru} by Forman et al
(1978). Since then it has been observed using {\sl EXOSAT} (Shrader et
al 1988), {\sl Ginga} (Ishida et al 1991, Done \& Magdziarz 1998),
{\sl ASCA} (Kallman et al 1996, Done \& Magdziarz 1998), {\sl ROSAT}
(Ramsay et al 1994, Mason et al 1998) and {\sl BBXRT} (Kallman et al
1993). These observations have shown a highly variable X-ray light
curve. 

This variability is best shown in the {\sl Ginga} data which was taken
over 3 days and showed a periodic modulation in the first half which
disappeared in the second half and was thereafter dominated by strong
flaring.  This is consistent with X-ray observations of one other
asynchronous polar RX J2115--58 which showed highly variable light
curves (Ramsay et al 2000). This variability was interpreted as the
accretion flow rotating around the magnetic field of the white dwarf
and at certain instances the flow accretes onto the opposite
hemisphere of the white dwarf resulting in a different shaped light
curve.

The hard X-ray spectra of BY Cam has shown evidence for a
multi-temperature nature and complex absorption (Done \& Magdziarz
1998). Using {\ros} data, Ramsay et al (1994) found that BY Cam was
the one system which did not show a soft/hard X-ray ratio which was
correlated with magnetic field strength. In this paper we present
X-ray and simultaneous optical/UV observations of BY Cam using {\sl
XMM-Newton}.

\section{Observations}

BY Cam was observed for a short duration using {\sl XMM-Newton} on 26
August 2001. We show in Table 1 the observation details in the various
instruments. The EPIC instruments (imaging detectors covering the
energy range 0.1--10keV with moderate spectra resolution) were
operated in small window mode so that the effects of pile-up would be
minimised. The RGS detectors (high resolution spectrographs operating
in the 0.3--2.0keV range: den Herder et al 2001) were configured in
the standard spectroscopy mode. For the OM (an optical/UV 30cm
telescope: Mason et al 2001), data in 2 UV filters were obtained
(UVW1: peak effective area at 2900 \AA\ and bandwidth 750 \AA\, UVW2:
2120 \AA\, 450 \AA) and one visual band (roughly the $V$ band, 5500
\AA\, 350 \AA).

The data were processed using the {\sl XMM-Newton} {\sl Science
Analysis Software} (SAS) v5.2. For the EPIC pn detector (Str\"{u}der
et al 2001), data were extracted using an aperture of 40$^{''}$ arc
sec centered on the source. Background data were extracted from a
source free region of the same CCD. For the EPIC MOS detectors (Turner
et al 2001) we extracted the background from source free regions of
other CCDs which were in full frame mode. The background data were
scaled and subtracted from the source data. The OM data were analysed
in a similar way using {\tt omfchain} (using the version due to be
released as part of the SAS v5.3). Data were corrected for background
subtraction and coincidence losses (Mason et al 2001).  Background
subtracted RGS spectra were derived using the SAS tool {\tt rgsproc}.

In extracting the EPIC pn spectra, we used only single pixel events
and used the response file epn\_sw20\_sY9\_thin.rmf. In the case of
the MOS data we used the response files
m[1-2]\_thin1v9q19t5r5\_all\_15.rsp.

\begin{table}
\begin{center}
\begin{tabular}{llrr}
\hline
Instrument& Mode & Filter & Duration\\
\hline
EPIC pn & small window & thin & 5063 sec\\
EPIC MOS & small window & thin & 5924 sec\\
EPIC RGS & & & 6406 sec\\
OM & image/fast & UVW1 & 1500 sec \\
OM & image/fast & V & 1500 sec \\
OM & image/fast & UVW2 & 2000 sec\\
\hline
\end{tabular}
\end{center}
\caption{The observation details of the various instruments on-board {\sl
XMM-Newton}.}
\end{table}

\section{Light Curves}

\subsection{General Features}

We show in Figure 1 the X-ray and optical/UV light curves. We are not
able to phase the data on a spin or orbital ephemeris because of the
uncertainty in the precise degree of asynchronism in this system. The
light curves show that the X-ray observations commenced when the
source was relatively faint (and showed a low soft/hard ratio). The
source then increased in intensity and showed a significant amount of
flaring activity towards the end of the observation (especially in the
soft X-ray band) and showed an increased soft/hard ratio. The increase
in X-ray flux is probably as a result of a second accretion region on
the white dwarf rotating into view. Based on the flux levels found in
\S 4.2, BY Cam was fainter by a factor of 5 and 3 compared to when it
was observed using {\sl ASCA} and {\sl Ginga} but comparable to its
{\sl RXTE} observation. Although these comparisons are complicated by
the fact that the {\sl XMM-Newton} observations cover less than half
of the binary orbital period, they do suggest that BY Cam was in a
relatively high rather than a low accretion state when observed using
{\sl XMM-Newton}.

The UVW1 data shows a high count rate when the X-ray flux was low: the
count rate (14.3 ct/s) corresponds to flux of $\sim6.4\times10^{-15}$
\ergscm \AA\ (based on OM observations of isolated white dwarfs). The
mean $V$ band magnitude is $\sim$15.7 ($\sim1.9\times10^{-15}$ \ergscm
\AA), reaching a peak of $\sim$15.5. Since the $V$ band data covered
the period when the source was increasing in X-ray brightness it is
likely that peak optical brightness in that orbital cycle was brighter
than $V\sim$15.5. Silber et al (1997) present an extensive series of
optical data on BY Cam. These show that when BY Cam is in a high state
it can show a considerable variation in optical brightness, although
typically around $V\sim$15 mag. Although the count rate in the UVW2
filter is low (1.9 ct/s) it corresponds to a flux of
$\sim7.5\times10^{-15}$ \ergscm \AA.

\begin{figure*}
\begin{center}
\setlength{\unitlength}{1cm}
\begin{picture}(15,10.5)
\put(-1,-0.2){\includegraphics{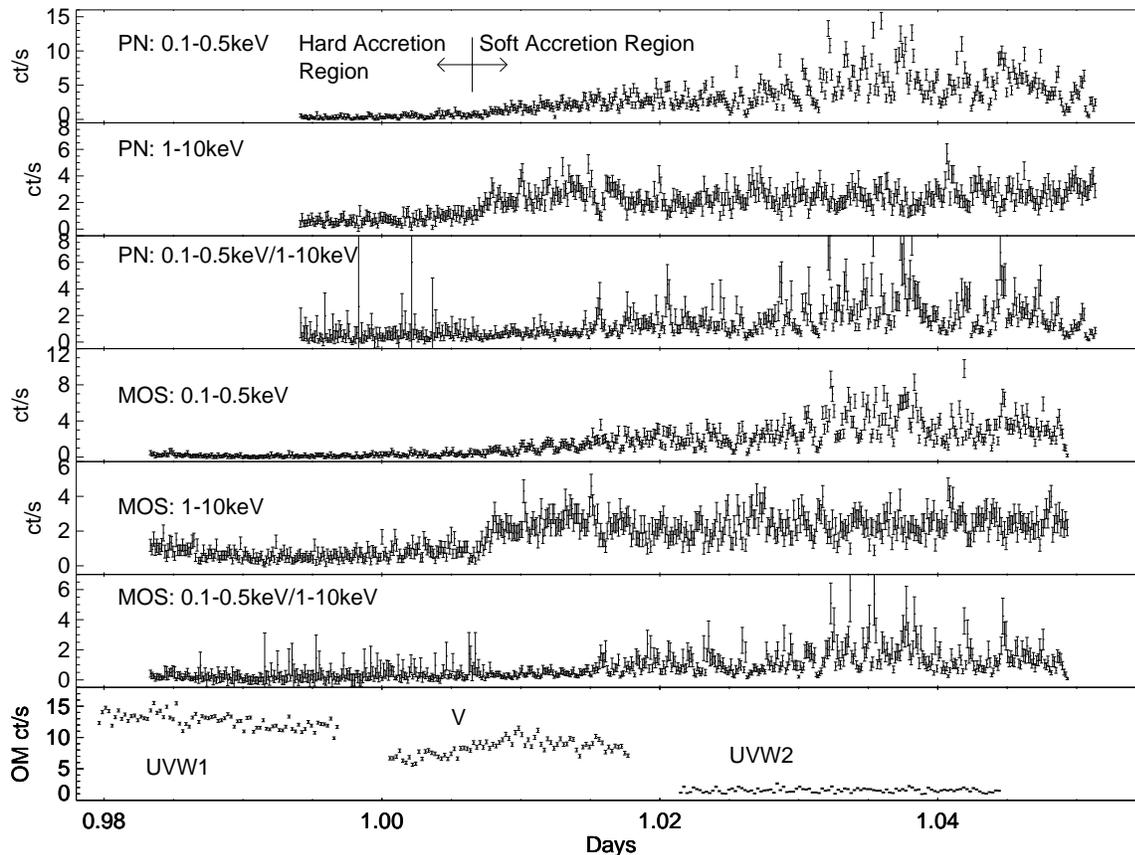}}
\end{picture}
\end{center}
\caption{The X-ray and optical/UV light curves of BY Cam observed
using {\sl XMM-Newton}. The MOS 1 and 2 data have been co-added.  The
days are TDB-2452147.0 corrected for light travel time to the solar
system barycentre. The `hard' and `soft' accretions denote time
intervals in the text.}
\label{light} 
\end{figure*}

\subsection{Short timescale variability}

The first half of the X-ray light curves are relatively stable showing
only small scale variability, mainly in the 1--10keV energy band (cf
Figure 1). Around day 1.025, flaring activity starts to become
prominent -- particularly at energies below 0.5keV. To study the
nature of the short term flaring in greater detail we examined data in
the time range starting from this point until the end of the
observation.

\begin{figure}
\begin{center}
\setlength{\unitlength}{1cm}
\begin{picture}(7,16)
\put(-2.5,-2){\includegraphics{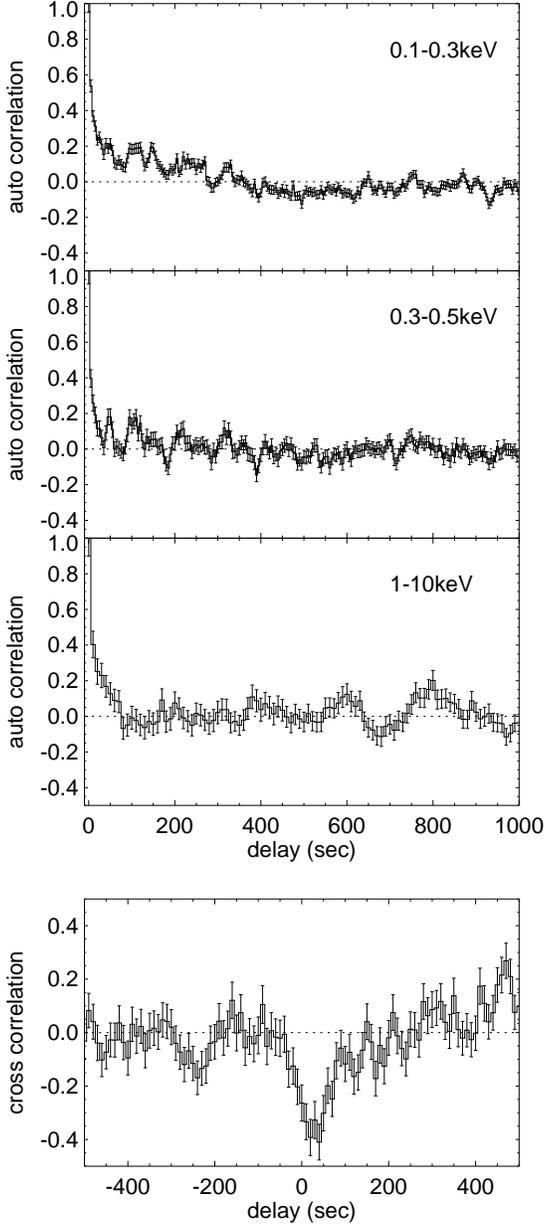}}
\end{picture}
\end{center}
\caption{From the top: The autocorrelation for the EPIC pn light curve
in the 0.1-0.3keV, 0.3--0.5keV and 1--10keV energy bands using data
from day 1.025 until the end (cf Figure 1). The time bin is 5 sec in
each case. Bottom panel: the cross
correlation function of the 0.1--0.3keV and 1--10keV light curves. The
positive time lag implies the hard X-ray light curve {\sl trails} the
soft X-ray light curve and is anti-correlated.}
\label{cross} 
\end{figure}

We show in Figure \ref{cross} the auto-correlation function of the
energy resolved light curves using the EPIC pn data. This shows that
the zero-crossing time in the 0.3--0.5keV and 1--10keV light curves
are both around $\sim$85 sec. However, in the softest band
(0.1--0.3keV), this value is much longer taking around $\sim$300
sec. In comparison, observations of the polar EF Eri showed the
zero-crossing time was $\sim$85 sec in {\sl EXOSAT} (both in the
CMA:0.04--2keV and ME: 2--6keV) observations (Watson, King \& Williams
1987) and around $\sim$55 sec in both soft and hard bands of {\ros}
data (Beuermann, Thomas \& Pietsch 1991).

We also cross-correlated the EPIC pn 0.1--0.3keV and 1--10keV light
curves (Figure \ref{cross}). This shows that the peak of the cross
correlation deviates from zero (25$\pm$15 sec) and is significantly
greater than other peaks in the delay range --400 to +400 sec. This
suggests that this correlation is significant and implies that the
1--10keV light curve {\sl trails} the 0.1--0.3keV light curve
and is {\sl anti-correlated}. This means that when
there is a brightening in the 0.1--0.3keV light curve there is a dip
in the 1--10keV light curve shortly afterwards. We also repeated the
same analysis using the EPIC MOS data and find the same result as for
the EPIC pn data. This is an entirely unexpected result which is
discussed further in \S \ref{discuss}.

We made a search for Quasi Periodic Oscillations (QPO) in the X-ray
light curve. This was done by examining a number of different time
intervals and energy bands. We found no evidence for a significant QPO
signal (on timescales between 0.1 sec and 50sec) in the Discrete Fourier
Transforms.

\section{X-ray spectra}

\subsection{RGS spectra}

\begin{figure}
\begin{center}
\setlength{\unitlength}{1cm}
\begin{picture}(7,6)
\put(-1.1,-0.5){\includegraphics{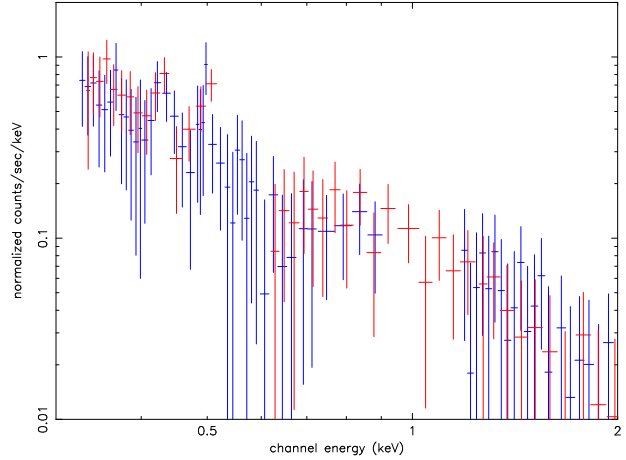}}
\end{picture}
\end{center}
\caption{The integrated spectrum in the RGS 1 \& 2
instruments. Although the signal to noise of the spectra are low,
lines are seen at 0.426keV and 0.500keV in both instruments.}
\label{rgsspec} 
\end{figure}

Although the count rate in the RGS instruments was low ($\sim$0.2
ct/s), it was sufficient to extract a spectrum with sufficient
signal-to-noise to search for prominent features. We show in Figure
\ref{rgsspec} the RGS1 and RGS2 spectra. There are two lines which are
apparent in both instruments suggesting they are real instruments: at
0.427$^{+0.011}_{-0.006}$keV and 0.500$^{+0.002}_{-0.003}$ keV. They
have equivalent widths of 8$^{+21}_{-2}$eV and 35$^{+31}_{-15}$eV
respectively. They are probably due to NVI (0.4262 or 0.4307 keV) and
NVII (0.500 keV).

\subsection{EPIC spectra}

The hardness ratios shown in Figure 1 suggest that the accretion
region which is viewed at the beginning of the observation is harder
than the second accretion region which come into view at day 1.007.
We therefore extracted spectra from time intervals corresponding to
the first visible accretion region (the `hard' region) and the second
visible region (the `soft' region). Our emission model consisted of
neutral absorbing material and a multi-temperature shock model
(Cropper et al 1999). This model takes into account cooling via
cyclotron radiation, a variation in the gravitational force over the
height of the post-shock flow and the multi-temperature nature of this
region: the immediate layers beneath the shock front reaches
temperatures $\sim$30-50keV, while closer to the white dwarf
photosphere it has cooled to $\sim$1keV. We fix the ratio of cyclotron
to bremsstrahlung cooling, $\epsilon_{s}$ at 5, appropriate for the
magnetic field strength of the white dwarf in BY Cam (thought to be in
the range 30--50MG: Cropper et al 1989). When necessary we added a
neutral absorption which had a partial covering fraction and a
blackbody. We also added a Gaussian component to model the fluorescent
line at 6.4keV. We fitted the spectra from the three EPIC detectors
separately.

\begin{figure*}
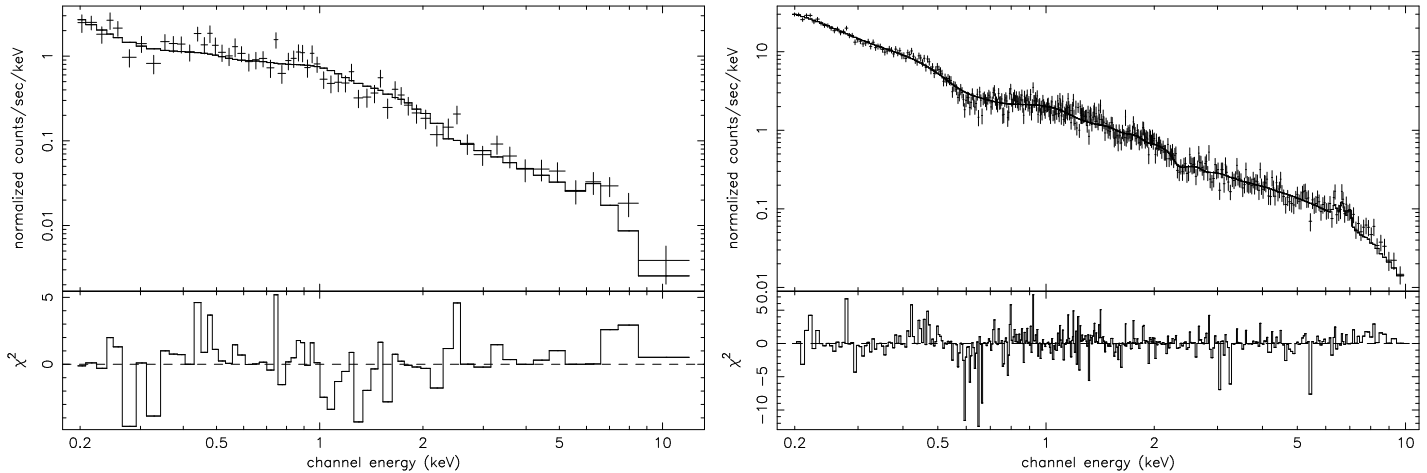

\begin{center}
\setlength{\unitlength}{1cm}
\begin{picture}(7,6)
\put(-6,-0.5){\includegraphics{spec_hard_region_chi.ps}}
\put(3.5,-0.5){\includegraphics{spec_soft_region_chi.ps}}
\end{picture}
\end{center}
\caption{The EPIC pn spectra of the `hard' (left hand panel) accretion
region and the `soft' accretion region (right hand panel). A soft
X-ray component is only seen in the `soft' accretion region.}
\label{spec} 
\end{figure*}

We show in the left hand panel of Figure \ref{spec} the EPIC pn
spectrum covering the hard time interval. The spectral parameters for
all three spectra are shown in Table \ref{spechard}. The immediate
points to note are the absence of a soft blackbody component and the
low column density. We define the hard X-ray luminosity as
$L_{hard}=4\pi$Flux$_{hard,bol}d^{2}$ where Flux$_{hard,bol}$ is the
unabsorbed, bolometric flux from the hard component and $d$ is the
distance. Since a fraction of this flux is directed towards the
observer, we switch the reflected component to zero after the final
fit to determine the intrinsic flux from the optically thin post-shock
region. For a distance of 100pc we find that $L_{hard}\sim10^{31}$
\ergss.

For the spectra covering the soft time interval we find that a soft
blackbody component is required to achieve good fits to the
data. Further, in addition to the neutral absorption component we
require a second absorption component with a partial covering
fraction. We show the EPIC pn spectrum in the right hand panel of
Figure \ref{spec} and the spectral parameters in Table \ref{specsoft}.
We define the soft X-ray luminosity as
$L_{soft}=\pi$Flux$_{soft,bol}$sec($i-\beta)d^{2}$, where we assume
that the soft X-ray emission is optically thick and can be
approximated by a small thin slab of material, the unabsorbed
bolometric flux is Flux$_{soft,bol}$, $i$ the inclination and $\beta$
the angle between the accretion region and the spin axis. Taking
$i\sim$40--60$^{\circ}$ and $\beta\sim$100--125$^{\circ}$ (assuming
the soft pole is negatively circularly polarised: Piirola et al 1994)
we find sec$(i-\beta)\sim$2. The ratio of
$L_{soft}/L_{hard}\sim$0.5. This is similar to that found by Ramsay et
al (1994) using {\sl ROSAT} data. They suggested that because BY Cam
is an asynchronous system the accretion region may have a larger area
than is typical in polars because the aspect of the white dwarf is
continuously changing with respect to the accretion stream. Extracting
a spectrum from the second half of the soft interval we find that the
soft/hard flux increases by a factor of $\sim$4: this is consistent
with the view that the soft flaring is due to dense blobs of material
which impact directly into the photosphere of the white dwarf without
forming a shock.

Our spectral fits also allow us to determine the mass of the white
dwarf using the Nauenberg (1972) mass-radius relationship for white
dwarfs. Tables \ref{spechard} and \ref{specsoft} show that we find a
mass 0.9--1.1\Msun. This is consistent with the mass determined using
{\sl Ginga} data (0.98\Msun: Cropper et al 1999) and {\sl RXTE} data
(1.04\Msun: Ramsay 2000).

\begin{table*}
\begin{center}
\begin{tabular}{llrrrrrr}
\hline 
Detector& $N_{H}$ & $M_{1}$ & Observed flux& Unabsorbed flux &
Shock flux & \rchi & $L_{X}$ \\ 
& ($\times10^{19}$ & (\Msun) &
(0.1--10keV) & Bolometric & Bolometric & (dof) & \ergss \\ & \pcmsq) &
& \ergscm & \ergscm & \ergscm & & \\ 
\hline 
EPIC pn & 0$^{+4}$ &
0.9$^{+0.2}_{-0.1}$ & 6.1$\pm0.5\times10^{-12}$ &
1.0$\pm0.1\times10^{-11}$ & 9.7$\pm0.9\times10^{-12}$ & 1.38 (57)
& 1.2$\pm0.1\times10^{31}$ \\ 
EPIC MOS1 & 0$^{+8}$ &
1.14$^{+0.12}_{-0.17}$ & 6.0$\pm1.5\times10^{-12}$ &
1.2$\pm0.3\times10^{-11}$ & 1.1$\pm0.3\times10^{-11}$ & 1.34 (49)
& 1.3$\pm0.4\times10^{31}$ \\ 
EPIC MOS2 & 0 & 0.94$^{+0.10}_{-0.16}$ &
5.8$^{+1.6}_{-0.6}\times10^{-12}$ &
1.0$^{+0.3}_{-0.1}\times10^{-11}$ &
9.5$^{+2.6}_{-1.0}\times10^{-12}$ & 0.99 (92) &
1.1$^{+0.4}_{-0.1}\times10^{31}$\\ 
\hline
\end{tabular}
\end{center}
\caption{The results of our spectral fits to the EPIC data covering
the hard accretion region. The errors levels refer to the 90 percent
confidence level. The shock flux refers to the flux originating solely
from the post-shock flow and excluding the hard X-rays reflected from
the white dwarf. We assume a distance of 100pc in determining
the luminosities.}
\label{spechard}
\end{table*}

\begin{table*}
\begin{center}
\begin{tabular}{llrrrrrr}
\hline 
Detector & $N_{H}$ & $N_{H}$ & cvf & $kT_{bb}$ & Observed bb flux
& Bolometric bb  & $L_{soft}$ \\
 & ($\times10^{20}$ & ($\times10^{20}$ & & (eV) &
(0.1--10keV) ($10^{-12}$ & flux ($10^{-11}$ &  ($10^{31}$\\
 & \pcmsq) & \pcmsq) & & & \ergscm) &\ergscm)  
& \ergss) \\
\hline
EPIC pn & 2.7$\pm$0.5 & 7$^{+9}_{-2}$ & 0.41$^{+0.08}_{-0.06}$ & 54$\pm$3
& 5.8$^{+1.5}_{-1.4}$ & 3.9$\pm1.0$ &
2.3$\pm0.6$ \\
EPIC MOS1 & 2.0$^{+4.0}_{-0.6}$ & 8$^{+16}_{-1}$ & 0.40$\pm0.06$ &
54$^{+3}_{-6}$ & 6.2$^{+4.6}_{-2.3}$ & 5.3$^{+2.4}_{-1.1}$ &
3.2$^{+1.5}_{-0.7}$ \\
EPIC MOS2 & 2.2$^{+1.0}_{-0.8}$ & 17$^{+8}_{-5}$ & 0.56$^{+0.07}_{-0.08}$ &
52$^{+3}_{-4}$ & 7.7$^{+6}_{-2}$ & 8.4$^{+6.5}_{-2.1}$ &
5.0$^{+4.0}_{-3.8}$ \\
\end{tabular}
\begin{tabular}{llrrrrrr}
\hline
Detector & $M_{1}$ & Observed hard flux & 
Bolometric hard flux & Shock Bolometric & $L_{hard}$ &
$L_{soft}/L_{hard}$ & \rchi\\
 &  (\Msun) & (0.1--10keV) ($10^{-11}$ & flux ($10^{-11}$ & flux ($10^{-11}$
& ($10^{31}$  
&  & (dof) \\
 & & \ergscm) &\ergscm) & \ergscm) & \ergss) & & \\
\hline
EPIC pn & 0.91$^{+0.07}_{-0.06}$ &
2.0$^{+0.2}_{-0.1}$ & 4.6$^{+0.5}_{-0.2}$ &
4.4$^{+0.5}_{-0.2}$ & 5.3$^{+0.6}_{-0.3}$ &
0.43$^{+0.15}_{-0.14}$ & 1.07 (418)\\
EPIC MOS1 & 1.10$^{+0.05}_{-0.06}$ & 2.2$\pm$0.02 &
6.1$^{+0.5}_{-0.6}$ & 5.7$^{+0.5}_{-0.6}$ & 6.9$^{+0.6}_{-0.7}$ &
0.46$^{+0.3}_{-0.13}$ & 1.58 (124)\\
EPIC MOS2 & 1.01$^{+0.09}_{-0.18}$ & 2.3$^{+0.05}_{-0.3}$ &
7.6$^{+1.7}_{-1.0}$ & 7.2$^{+1.7}_{-1.0}$ & 8.6$^{+2.0}_{-1.2}$ &
0.58$^{+0.63}_{-0.47}$ & 1.15 (127)\\
\hline
\end{tabular}
\end{center}
\caption{The fit parameters to the `soft' accretion region. In the top
panel we show the flux from the blackbody component and the hard X-ray
component in the bottom. We assume a distance of 100pc in determining
the luminosities.}
\label{specsoft}
\end{table*}

\section{Discussion}
\label{discuss}

\subsection{Short time variability}

Various groups have searched for a correlation between the soft and
hard X-ray light curves of polars. Observations of AM Her (Stella,
Beuermann \& Patterson 1986) showed no correlation between the soft
and hard X-ray light curves while observations of EF Eri (Watson, King
\& Williams 1987) found that the soft and hard X-ray light curves were
weakly correlated at zero lag. However, in both of these cases, the
`soft' X-ray light curve extended up to 2keV or beyond and hence their
`soft' band may have been contaminated by photons originating from the
post-shock flow. With the good spectral resolution of {\xmm} we have
been able to separate the soft re-processed X-ray emission from the
post-shock emission.

In \S 3.2 we found evidence that the hard X-ray light curve trailed
the softest X-ray light curve by $\sim$10--40sec and was
anti-correlated. This is an unexpected result. The emission at
energies 0.1--0.3keV is produced by a combination of hard X-rays
irradiating the white dwarf and then being re-radiated at lower
energies (Lamb \& Masters 1979, King \& Lasota 1979), and dense blobs
of material penetrating the photosphere of the white dwarf (Kuijpers
\& Pringle 1982). The proportion of blobs present in the accretion
flow is thought to be proportional to the magnetic field strength of
the white dwarf (Ramsay et al 1994). At energies above $\sim$1keV the
emission is due entirely from X-rays produced in the post-shock region
above the photosphere of the white dwarf. If soft X-rays were produced
largely from the re-processing of hard X-rays then they would be
expected to be correlated (it is possible there would be a short time
delay with the soft band trailing the hard since it would take a
finite time for the hard X-rays to be reprocessed).  We suggest that
the observed delayed anti-correlation is due to dense blobs of
material impacting the photosphere of the white dwarf which then
throw-up enough optically thick material to obscure even hard X-rays
for a short time. Supporting evidence is found from the spectral fits
which indicate the absorption is significantly higher in the soft
accretion region spectrum compared to the hard accretion region
spectrum.

\subsection{Spectra}

The RGS spectra show the presence of line emission at 0.426 and 0.500
keV. These lines appear to be due to NVI and NVII. We believe that
this is the first time that Nitrogen lines have been seen in the X-ray
spectra of a CV. This observation is interesting since Bonnet-Biduad
\& Mouchet (1987) using {\sl IUE} spectra found that the ratio of
NV/CIV is unusually large - around a factor of 20 greater than
normally found in polars. They suggested that this unusual ratio was
due to a previous nova event in the system. The lines detected in the
X-ray band are of course more ionised that those seen in the UV but
their origin (whether in the pre-shock accretion flow, circumbinary
material or from the white dwarf photosphere), is still unclear.

Another significant result from our spectral fitting is the finding
that no blackbody component is seen in the `hard' accretion region
while it is in the `soft' accretion region. A soft X-ray component has
been seen in many polars, but there are examples where none have been
detected. The polar WW Hor was observed using {\xmm} (Ramsay et al
2001) in an intermediate accretion state and no soft X-ray component
was detected. Further, the soft and hard X-ray light curves are almost
always seen in phase. However, there have been some observations where
the soft and hard X-ray components have been seen in anti-phase (the
`reversed X-ray mode'). One example is the observation of AM Her made
using {\sl EXOSAT} (Heise et al 1985). This observation contrasts with
other observations of AM Her which show the soft and hard light curves
in phase.

Heise et al (1985) suggest that AM Her normally accretes only onto one
magnetic pole, but in the reversed X-ray mode, accretion occurs onto
both poles. When accreting onto only one pole (the primary pole), both
soft and hard X-rays are generated. When accretion occurs on the
secondary pole, accretion takes place in the form of dense blobs of
material which penetrate directly the white dwarf photosphere and no
shock is formed: this pole is therefore strong in soft X-rays.
Polarimetric observations of BY Cam by Piirola et al (1994) show
evidence for positive and negative circular polarisation, implying the
presence of two accretion poles.

Dealing with the `soft' pole first, there is clear evidence of short
term variations in the soft X-ray light curve (at least in the second
half of the soft light curve). Since there are no short term
variations in the hard X-ray light curve, the most probable mechanism
for the soft flaring is dense blobs of material which impact directly
into the white dwarf. This would be expected to reflected in a high
$L_{soft}/L_{hard}$ ratio. In contrast, the fits to the spectrum of
the soft accretion region imply a ratio $\sim$0.5 which is consistent
with the `standard' accretion model in which the soft X-ray emission
originates entirely from re-processing of hard X-rays. However, when
we consider the spectrum taken from the only the flaring state, we
find that the soft/hard ratio increases by a factor of 4. This is
consistent with the view that dense blobs of material are indeed
present in the flow and are the cause of flaring in soft X-rays.

We now discuss the hard accretion region. Compared to the soft
accretion region, the hard region shows a light curve which is much
more constant in intensity. This suggests the absence of dense blobs
of material impacting onto this accretion pole. However, we would
still expect there to be a soft X-ray component from re-processing of
the hard X-ray component. If the accretion rate was reduced (the hard
X-ray flux is a factor $\sim$5 lower in this region compared to the
soft region) or the flow was spread over a larger area than usual,
then the temperature of the re-processed component would be
lower. This would result in the re-processed component being shifted
into UV wavebands. Indeed, Figure 1 shows that the UVW1 count rate is
high when the hard region is in view.

To investigate this further, we re-considered the soft accretion
region spectrum by adding a blackbody of temperature 1.7eV (20000K,
G\"{a}nsicke, Beuermann \& de Martino 1998) to account for that part
of the white dwarf not heated by the accretion flow. We then set the
normalisation of this component to match the flux observed in the UVW2
filter. We then added this low temperature blackbody to our model for
the hard accretion region.  We find that the observed flux in the UVW1
filter is higher by a factor $\sim$2 than predicted. We now add a
second blackbody to account for a re-processed component: this is
fixed at various temperatures and normalisations so that it gives the
observed UVW1 flux and has no significant effect on the fit to the
X-ray spectra.

We find that for a heated region of temperature $\sim$4eV (as opposed
to 54eV in the soft accretion region), we find
$L_{soft}/L_{hard}\sim$1. (Here we set sec($i-\beta$)=1 appropriate
for the positively circularly polarised pole which we take to be the
hard accretion region, Piirola et al 1994). Some degree of caution
should be applied there are uncertainties in the exact temperature of
the unheated white dwarf, the precise UV flux estimates and the amount
of absorption. However, this supports our view that the lack of a
distinct soft X-ray component in the hard accretion region is because
this component has moved into the UV as a result of the lower hard
X-ray heating rate, or that it is spread over a larger area compared
to the soft accretion pole. This is similar to that observed in AM Her
(G\"{a}nsicke, Beuermann \& de Martino 1998).

\end{document}